\documentclass{article}%
\usepackage{amsmath}
\usepackage{amsfonts}
\usepackage{amssymb}
\usepackage{graphicx}
\usepackage{slashed}%
\providecommand{\U}[1]{\protect\rule{.1in}{.1in}}
\newtheorem{theorem}{Theorem}
\newtheorem{acknowledgement}[theorem]{Acknowledgement}

\begin{document}
\begin{titlepage}
\vspace{.3cm} \vspace{1cm}
\begin{center}
\baselineskip=16pt \centerline{\Large\bf  Noncommutative Geometry and Structure of Space-Time\footnote{Plenary Talk
		at $9^{th}$ Pan African Congress of Mathematicians, PACOM 2017, Morocco, to appear in Special Issue of Africa Matematica, Editor J. Banasiak}}
\vspace{.5truecm}
\vspace{.5truecm}
\centerline{\large\bf Ali H.Chamseddine$^{1,2}$ } \vspace{.5truecm}
\emph{\centerline{$^{1}$Physics Department, American University of Beirut, Lebanon}}
\emph{\centerline{$^{2}$I.H.E.S. F-91440 Bures-sur-Yvette, France}}
\end{center}
\vspace{2cm}
\begin{center}
{\bf Abstract}
\end{center}
I give a summary review of the research program using noncommutative geometry
as a framework to determine the structure of space-time. Classification of finite
noncommutative spaces under few assumptions reveals why nature chose the Standard
Model and the reasons behind the particular form of gauge fields, Higgs fields and
fermions as well as the origin of symmetry breaking. It also points that at high energies the
Standard Model is a truncation of Pati-Salam unified model of leptons and quarks.
The same conclusions are arrived at uniquely without making any assumptions except
for an axiom which is a higher form of Heisenberg commutation relations quantizing
the volume of space-time. We establish the existence of two kinds of quanta of
geometry in the form of unit spheres of Planck length. We provide answers to many of the
questions which are not answered by  other approaches, however, more research is needed
to answer the remaining challenging questions.
\end{titlepage}

\section{Introduction}

One of the main aims of theoretical physics is to explain and understand the
fundamental laws of nature. Despite the apparent complexity, the laws of
physics are expressed by concise mathematical formulas. Presently there are
four known fundamental forces. The first force is gravity which is universal.
It interacts with all particles with mass or energy. It is extremely weak and
long range and mediated by the graviton, a tensor $g_{\mu\nu}$. Due to the
absence of negative mass and the attractive nature of gravity, masses can
become extremely large and the gravitational force sizable as in the examples
of stars and black holes. The electromagnetic force acts on charged particles,
mediated by the photon $A_{\mu},$ where charges could be positive or negative
and is much stronger ($\sim10^{36}$) than the gravitational force. Despite the
huge ratio, astronomical objects are mostly neutral, resulting in the
dominance of gravity. The third force is the weak force covering the
interactions of doublets and singlets in left-handed and right-handed sectors
of leptons and quarks. The electromagnetic and weak interactions are unified
in one force, the electroweak force at the energy scale of $\sim100$ Gev
mediated by the four vector bosons of the symmetry group $SU\left(  2\right)
\times U\left(  1\right)  .$ This symmetry is broken spontaneously when the
scalar Higgs $SU\left(  2\right)  $ doublet acquires a vacuum expectation
value (vev) with the electromagnetic symmetry $U\left(  1\right)
_{\mathrm{em}}$ as the unbroken subgroup. The three vectors $W_{\mu}^{\pm}$
and $Z_{\mu}$ associated with the broken generators are heavy with masses of
$\sim80$ Gev and $\sim90$ Gev, respectively, thus explaining the short range
of weak interactions. The strong nuclear force acts on quarks made of three
different colors and is thus governed by the symmetry group $SU(3).$ The force
is mediated by \ the eight Gluon vectors, $V_{\mu}^{i},$ $i=1,\cdots,8$ and is
$\sim10^{2}$ stronger than the electromagnetic force. The fundamental fermions
come in three identical families in representations of the symmetry group
$SU\left(  3\right)  \times SU\left(  2\right)  \times U\left(  1\right)  $
where each member of the higher generation have a much larger mass
($\sim10^{2}$) than the corresponding particle of the previous generation.
There are $16$ fermions per generation, each comprising of $8$ left-handed and
$8$ (anti) right-handed two components chiral spinors. All masses are acquired
through the couplings of the Higgs doublet to the left and right handed
spinors. The $8$ left-handed components of the first family comprise of a
lepton $SU\left(  2\right)  $ doublet made from a neutrino and an electron as
well as a quark doublet made from up and down quarks of three colors. The $8$
right-handed components comprise of $SU\left(  2\right)  $ singlets made up of
a neutrino, electron, an up quark and down quark of three colors. The
neutrinos (which are electrically neutral)\ are mostly left-handed with
extremely small masses which is explained by assuming that the right-handed
components also acquire a Majorana type large mass, usually through coupling
to a singlet scalar field with a large vev $\left(  \sim10^{11}\text{ Gev
}\right)  .$ The particles listed above with the forces governing their
interactions (without gravity) is known as the Standard Model of particle
physics. The model has $19$ parameters, fixed by experimental data, and is
studied using the methods of quantum field theory. It agrees, so far, with all
experimental tests and at present there are no hints of whether there is
physics beyond the Standard Model. There are, in addition, parameters
associated with the neutrinos, which are still not fully determined from
experiment. The success of the Standard Model poses enormous challenge to
understanding the reasons for its very specific structure. Out of infinite
number of possibilities the question we must pose is to understand the reason
why nature chose the configuration given by the Standard Model. One important
question we will answer in this lecture is \textit{Why the Standard Model? }

The monumental achievement of Einstein is that he showed that matter curves
the four-dimensional space-time in such a way that the geometry becomes
Riemannian. The main tool in Riemannian geometry is the four-dimensional
metric tensor $g_{\mu\nu},$ $\mu,\nu=0,1,2,3,4$ which is used in measuring
distances
\[
ds^{2}=g_{\mu\nu}dx^{\mu}dx^{\nu}%
\]
A test particle in this space  moves along a geodesic according to the
parametric equation%
\[
\frac{d^{2}x^{\mu}}{d\lambda^{2}}+\Gamma_{\nu\rho}^{\mu}\frac{dx^{\nu}%
}{d\lambda}\frac{dx^{\rho}}{d\lambda}=0
\]
where $\lambda$ is the affine parameter and $\Gamma_{\nu\rho}^{\mu}$ is the
(symmetric) Christoffel connection determined from the metricity condition
\[
0=\nabla_{\rho}g_{\mu\nu}=\partial_{\rho}g_{\mu\nu}-\Gamma_{\rho\mu}^{\sigma
}g_{\sigma\nu}-\Gamma_{\rho\nu}^{\sigma}g_{\mu\sigma}%
\]
Dynamics of the gravitational field is dictated by the Einstein-Hilbert
action
\[
I_{g}=-\frac{1}{16\pi G}%
{\displaystyle\int}
d^{4}x\sqrt{g}R
\]
where $R$ is the scalar curvature of the four-dimensional manifold and $G$ is
Newton constant. The variational principle gives Einstein equations, which is
a set of $10$ non-linear second order differential equations
\[
G_{\mu\nu}\equiv R_{\mu\nu}-\frac{1}{2}g_{\mu\nu}R=8\pi GT_{\mu\nu}%
\]
where $G_{\mu\nu}$ is the Einstein tensor and $T_{\mu\nu}$ is the
energy-momentum tensor due to the contributions of matter. Newton's law is
recovered by setting $g_{00}=1+\phi$ and $g_{ij}=-\delta_{ij}$ where $\phi$ is
the gravitational potential. This classical action has passed all experimental
tests including the recent results on gravitational waves. This action is,
however, not well behaved at the quantum level. At present, there is no
satisfactory theory of quantum gravity. In this sense gravity cannot be
unified with the other three fundamental interactions. \ A main difficulty is
to quantize the metric $g_{\mu\nu}$ in a background independent way. On the
other hand, a very important question to answer is whether there is a scale at
which all the four interactions are unified? It is known, to a high degree of
precision that if there is no new physics beyond the Standard Model, then the
three coupling constants, which are energy dependent, do not all meet at one
common energy scale. Instead of intersecting at a point, the intersection of
the three lines (on a logarithmic scale)  form a triangle in the energy range
$10^{14}-10^{17}$ Gev, which is not far off from the Planck scale $10^{19}$
Gev where the gravitational force becomes strong. For a believer in the idea
of unification, this suggests that there must be new physics beyond the
Standard Model that makes unification a possibility and it is preferable to
have the unification scale to be near the upper end of $10^{17}$ Gev. This
would also point to the possibility of having all the fields, including the
graviton, as part of a full unified system and not as separate independent
parts. In addition, the possibility of unification with gravity at the Planck
scale where the gravitational field becomes strong, suggests that Riemannian
geometry would cease to be the appropriate framework to describe geometry of
space-time. Some discretization must take place, and there must exist a
minimal volume, or some form of quanta of geometry, out of which space-time is built.

The purpose of this summary is to show that it is possible to understand the
deep reasons behind the choice of nature to the Standard Model and why the
particles, their interactions and group representations are the way they are.
The framework of Riemannian geometry must be replaced by that of
noncommutative geometry combined with the condition of the existence of quanta
of geometry. Noncommutative geometry is spectral in nature. It grew from
formulation of quantum mechanics by von Neumann and later developed by Alain
Connes \cite{Connesbook}. The geometric data is given by a spectral triple
$\left(  \mathcal{A},\mathcal{H},D\right)  $ where $\mathcal{A}$ is an
associative algebra with unit $1$ and involution $\ast$, $\mathcal{H}$ a
complex Hilbert space carrying a faithful representation of the algebra
$\mathcal{A}$ and $D$ is a slef-adjoint operator on $\mathcal{H}$ with a
compact resolvent of $\left(  D-\lambda1\right)  ^{-1},$ where $\lambda
\notin\mathbb{R}$. In addition the real structure $J$ \ is an anti-unitary
operator that sends the algebra $\mathcal{A}$ to its commutant $\mathcal{A}%
^{o}$ such that \cite{Reality}
\begin{equation}
\left[  a,b^{o}\right]  =0,\qquad a,b\in\mathcal{A},\qquad b^{o}=Jb^{\ast
}J^{-1}\in\mathcal{A}^{o}.
\end{equation}
The chirality operator $\gamma$ is a unitary operator in $\mathcal{H}$ defined
in even dimensions such that $\gamma^{2}=1$ and commutes with $\mathcal{A}$
\begin{equation}
\left[  \gamma,a\right]  =0\qquad\forall a\in\mathcal{A}.
\end{equation}
There are commutativity or anti-commutativity relations between $D,$ $J,$ and
$\gamma:$%
\begin{equation}
J^{2}=\epsilon,\qquad JD=\epsilon^{\prime}DJ,\qquad J\gamma=\epsilon
^{\prime\prime}\gamma J,\qquad D\gamma=-\gamma D,
\end{equation}
where $\epsilon,\epsilon^{\prime},\epsilon^{\prime\prime}\in\left\{
-1,1\right\}  .$ The operators $\gamma$ and $J$ are similar to the chirality
and charge conjugation operators and to every fixed value of $\epsilon
,\epsilon^{\prime},\epsilon^{\prime\prime}$ is associated a KO dimension,
which may be non-metric, and thus is defined only modulo $8.$ Riemann's
formula for the distance between two points
\[
d\left(  a,b\right)  =\mathrm{\inf}%
{\displaystyle\int\limits_{\gamma}}
\sqrt{g_{\mu\nu}dx^{\mu}dx^{\nu}}%
\]
where $\gamma$ is any path connecting $a$ and $b,$ is replaced by
\[
d\left(  a,b\right)  =\sup\left\{  \left\vert f\left(  a\right)  -f\left(
b\right)  \right\vert ;\quad f\in\mathcal{A},\quad\left\Vert \left[
D,f\right]  \right\Vert \leq1\right\}
\]
It is useful to get familiarized with the new geometry by noting that
Riemannian geometry corresponds to the choice where $\mathcal{A}=C^{\infty
}\left(  M\right)  $ the algebra of infinitely differentiable functions over
$M,$ $\mathcal{H}=L^{2}\left(  S,M\right)  ,$ the Hilbert space of square
integrable functions and
\[
D=e_{a}^{\mu}\gamma^{a}\left(  \partial_{\mu}+\omega_{\mu}\right)
\]
where $e_{a}^{\mu}$ is the square root of the inverse metric $g^{\mu\nu},$
$\gamma^{a}$ are anti-Hermitian Dirac matrices defining a Clifford algebra
$\left\{  \gamma^{a},\gamma^{b}\right\}  =-2\delta^{ab},$ $a,b=1,\cdots,d$
where $d$ is the dimension of $M$. The spin-connection $\omega_{\mu}$ is
$SO\left(  d\right)  $ Lie algebra valued, $\omega_{\mu}=\frac{1}{4}%
\omega_{\mu}^{bc}\gamma_{bc},$ where $\gamma_{bc}=\frac{1}{2}\left(
\gamma_{b}\gamma_{c}-\gamma_{c}\gamma_{b}\right)  .$ The spectrum of the
operator $D$ contains all the geometric invariants. In simpler terms, it is
more useful to use spectrum of the Dirac operator, which we interpret as the
inverse distance: $ds^{-1}=D,$ to measure distances. The reality operator $J$
is the charge conjugation operator and $\gamma$, the chirality operator in
even dimensions, is
\[
\gamma=\left(  i\right)  ^{\frac{d}{2}}\gamma_{1}\cdots\gamma_{d}%
\]
Given the above identifications, results of Riemannian geometry over
spin-manifolds are obtained with the aid of reconstruction theorems
\cite{Gracia}. 

\section{Classification and Why the Standard Model}

Marc Kac \cite{Kac} asked the question: \textit{Can we hear the shape of the
drum? }Geometrically this is equivalent to determining the geometry of a
manifold from the spectrum of the Laplacian. Although the existence of
isospectral manifolds showed that the answer is not unique \cite{Milnor}, this
deficiency is overcome in noncommutative geometry where the data defines
uniquely the geometry. The question that Kac asked is relevant to determining
the structure of space-time as defined by the spectrum of all the known
elementary particles. Naturally our notion of space-time will get modified
whenever more particles at higher energies, or equivalently small distances,
are discovered. As alluded to in the introduction, there is no known
connection between the elementary particles of the Standard Model and
geometry. Although gauge theories can be thought of as vector bundles defined
over the four-dimensional manifold, this description does not add any
understanding to answer the question: \textit{Why the Standard Model?
}Fortunately the task of identifying the structure of space-time, in the down
to up approach as function of energy, starting with the spectrum of known
particles, could be answered in noncommutative geometry \cite{AC2}
\cite{saddam}. This shows the need for a geometry that combines the
four-dimensional Riemannian manifold which has continuos spectrum with the
discrete spectrum of the Standard Model. We must therefore look for a
noncommutative space admitting both discrete and continuous spectrum
\cite{Gros}. This situation is reminiscent of quantum mechanics, the
inspiration behind noncommutative geometry, where physical operators such as
the Hamiltonian, admit both discrete and continuos spectrum . As a first
approximation we assume that the required noncommutative geometry is a tensor
product of two geometries, the first corresponding to a continuous
four-dimensional Riemannian spin-manifold, and the second is discrete. The
spectrum of the full geometry is then given by \cite{AC2}
\begin{align*}
\mathcal{A} &  =C^{\infty}\left(  M_{4}\right)  \otimes\mathcal{A}_{F}\\
\mathcal{H} &  =L^{2}\left(  S,M_{4}\right)  \otimes\mathcal{H}_{F}\\
D &  =D_{M}\otimes1+\gamma_{5}\otimes D_{F}\\
J &  =J_{M}\gamma_{5}\otimes J_{F}\\
\gamma &  =\gamma_{5}\otimes\gamma_{F}%
\end{align*}
where $\left(  \mathcal{A}_{F},\mathcal{H}_{F},D_{F},J_{F},\gamma_{F}\right)
$ defines the geometry of the finite space, which to start with, is
undetermined. Irreducible elements of the Hilbert space $\Psi\in\mathcal{H}$
are chiral $\gamma\Psi=\Psi,$ however, acting on them by the reality operator
$J$ produce new independent elements. Physically speaking, this implies that
all known fermions must have a mirror image with the same masses and this is
not observed. Thus we must require the property that $J\Psi$ will be
identified with $\Psi.$ This turns out to be satisfied only when the $KO$
dimension of the finite noncommutative space is $6$ (mod $8$). Classifying all
finite noncommutative spaces with $KO$ dimension $6$ shows that the algebra
$\mathcal{A}_{F}=M_{4n}\left(  \mathbb{C}\right)  \oplus M_{4n}\left(
\mathbb{C}\right)  ,$ $n\geq1$ \cite{CC2}. Making the further physical
requirement that the first algebras $M_{2n}\left(  \mathbb{C}\right)  $ is
subject to antilinear isometry symmetry, restricts it to the form
$M_{2n}\left(  \mathbb{H}\right)  $ where $\mathbb{H}$ is the field of
quaternions. The simplest possibility is then to have the algebra
$\mathcal{A}_{F}=M_{2}\left(  \mathbb{H}\right)  \oplus M_{4}\left(
\mathbb{C}\right)  $ and when this is subjected to the condition of
commutativity of elements of the algebra with the chirality operator
$\gamma_{F}$, the algebra reduces to $\mathcal{A}_{F}=\left(  \mathbb{H}%
_{R}\mathbb{\oplus H}_{L}\right)  \oplus M_{4}\left(  \mathbb{C}\right)  .$
Thus an element $\Psi\in\mathcal{H}$ is of the form $\Psi=\left(
\begin{array}
[c]{c}%
\psi_{A}\\
\psi_{A^{\prime}}%
\end{array}
\right)  $ where $\psi_{A}$ is a $16$ component $L^{2}\left(  M,S\right)  $
spinor in the fundamental representation of $\mathcal{A}_{F}$ of the form
$\psi_{A}=\psi_{\alpha I}$ where $\alpha=1,\cdots,4$ with respect to
$M_{2}\left(  \mathbb{H}\right)  $ and $I=1,\cdots,4$ with respect to
$M_{4}\left(  \mathbb{C}\right)  $ and where $\psi_{A^{\prime}}=C\psi
_{A}^{\ast}$ is the charge conjugate spinor to $\psi_{A}$ \cite{AC}. The
chirality operator $\gamma$ must commute with elements of $\mathcal{A}$ which
implies that $\gamma_{F}$ must commute with elements in $\mathcal{A}_{F}.$
This predicts the number of fundamental fermions to be $4^{2}$ chiral spinors
in the representations $\left(  2_{R},1_{L},4\right)  +\left(  1_{R}%
,2_{L},4\right)  $ of $SU\left(  2\right)  _{R}\times SU\left(  2\right)
_{L}\times SU\left(  4\right)  $ symmetry. At this point it is important to
note that the assumptions made are minimal and weak, except for one ad hoc
assumption of  the existence of antilinear isometry. Later we will show that
working with a completely different starting point, in the form of an axiom,
we arrive uniquely at the above algebra.

The operator $J$ sends the algebra $\mathcal{A}$ to its commutant, and thus
the full algebra acting on the Hilbert space $\mathcal{H}$ is
$\mathcal{A\otimes A}^{o}.$ Under automorphisms of the algebra
\begin{equation}
\Psi\rightarrow U\Psi,
\end{equation}
where $U=u\widehat{u}$ with $u\in\mathcal{A},$ $\widehat{u}\in\mathcal{A}^{o}$
with $\left[  u,\widehat{u}\right]  =0$, it is clear that the Dirac action is
not invariant. This is similar to the situation in electrodynamics where the
Dirac action is not invariant under local phase transformations but the
invariance is easily restored by introducing the vector potential $A_{\mu}$
through the transformation $\gamma^{\mu}\partial_{\mu}\rightarrow\gamma^{\mu
}\left(  \partial_{\mu}+ieA_{\mu}\right)  .$ In our case, the Dirac operator
$D$ is replaced with $D_{A}=D+A,$ where the connection $A$ is given by
\cite{CCS}
\begin{equation}
A=%
{\displaystyle\sum}
a\widehat{a}\left[  D,b\widehat{b}\right]  .
\end{equation}
It can be shown that under automorphisms $U$ of the algebra we have
$D_{A}\rightarrow UD_{A}U^{\ast}$ . The important special case occurs \ when
the connection $A$ belongs to the algebra $\mathcal{A}$ but not to its
commutant, and this occurs when the order one condition is satisfied
\begin{equation}
\left[  a,\left[  D,\widehat{b}\right]  \right]  =0,
\end{equation}
We then have
\[
A=A=A_{\left(  1\right)  }+JA_{(1)}J^{-1}%
\]
where $A_{\left(  1\right)  }=%
{\displaystyle\sum}
a\left[  D,b\right]  $ and the center of the algebra $Z\left(  \mathcal{A}%
\right)  $ is non-trivial in such a way that the space is connected. This
means that there is a mixing term between the fermions and their conjugates.
The Dirac operator connects the spinors $\psi_{A}$ and their conjugates
$\psi_{A^{\prime}}$ so that \cite{saddam}
\begin{equation}
\left[  D,Z\left(  \mathcal{A}\right)  \right]  \neq0.
\end{equation}
In physical terms this would allow a Majorana mass term for the fermions. It
was shown in \cite{CC2} that the unique solution to this equation constrains
the algebra $\mathcal{A}_{F}=\mathbb{H}_{R}\oplus\mathbb{H}_{L}\oplus
M_{4}\left(  \mathbb{C}\right)  $ to be restricted to a subalgebra
\cite{saddam}
\begin{equation}
\mathbb{C}\oplus\mathbb{H}_{L}\oplus M_{3}\left(  \mathbb{C}\right)  ,
\end{equation}
so that an element of $\mathcal{A}$ takes the form \cite{AC}
\begin{equation}
a=\left(
\begin{array}
[c]{ccccc}%
X\otimes1_{4} &  &  &  & \\
& \overline{X}\otimes1_{4} &  &  & \\
&  & q\otimes1_{4} &  & \\
&  &  & 1_{4}\otimes X & \\
&  &  &  & 1_{4}\otimes m
\end{array}
\right)  ,\qquad.
\end{equation}
where $X\in\mathbb{C},\quad q\in\mathbb{H},\quad$ $m\in M_{3}\left(
\mathbb{C}\right)  $ and the operator $D_{F}$ have a singlet non-zero entry in
the mixing term $\left(  D_{F}\right)  _{A}^{A^{\prime}}.$ The fermions are
enumerated as
\begin{align}
\psi_{\overset{.}{1}1} &  =\nu_{R},\quad\psi_{\overset{.}{2}1}=e_{R}\\
\psi_{a1} &  =l_{a}=\left(
\begin{array}
[c]{c}%
\nu_{L}\\
e_{L}%
\end{array}
\right)  \\
\psi_{\overset{.}{1}i} &  =u_{iR},\quad\psi_{\overset{.}{2}i}=d_{iR}\\
\psi_{ai} &  =q_{ia}=\left(
\begin{array}
[c]{c}%
u_{iL}\\
d_{iL}%
\end{array}
\right)  .
\end{align}
It is clear that the associated gauge group is $U\left(  1\right)  \times
SU\left(  2\right)  \times SU\left(  3\right)  .$ The (finite) Dirac operator
can be written in matrix form%
\begin{equation}
D_{F}=\left(
\begin{array}
[c]{cc}%
D_{A}^{B} & D_{A}^{B^{^{\prime}}}\\
D_{A^{^{\prime}}}^{B} & D_{A^{^{\prime}}}^{B^{^{\prime}}}%
\end{array}
\right)  ,\label{eq:dirac}%
\end{equation}
and must satisfy the properties
\begin{equation}
\gamma_{F}D_{F}=-D_{F}\gamma_{F}\qquad J_{F}D_{F}=D_{F}J_{F},
\end{equation}
where $J_{F}^{2}=1.$ We can show using elementary algebra, that the structure
of the connection $A$ is such that the block diagonal elements are of the form
$\gamma^{\mu}$ tensored with the gauge vectors of $U\left(  1\right)  \times
SU\left(  2\right)  \times SU\left(  3\right)  $ while the off-diagonal block
elements are of the form $\gamma_{5}$ tensored with the Higgs doublet complex
scalar field. There is, in addition, one singlet scalar field $\sigma$ (real,
or complex) \cite{Reselience}, \cite{Karimi}, not present in the Standard
Model whose vev gives mass to the right-handed neutrinos \cite{AC}. Dynamics
of all the fermion fields is then governed by the Dirac action \cite{saddam}
\begin{equation}
\left(  J\Psi,D_{A}\Psi\right)  .
\end{equation}
reproducing all the complicated structure present in the Standard Model. At
present, there is no known explanation for the three generations that we must
assume for the fermions and for the form of the Yukawa couplings responsible
for the hierarchy of fermion masses.

We have mentioned before that the spectrum of the Dirac operator is given by
geometric invariants. This led to the proposal that the dynamics of all
bosonic fields, including graviton, gauge fields and Higgs fields is governed
by the spectral action given by \cite{AC2}
\begin{equation}
\mathrm{Tr}f\left(  \frac{D_{A}}{\Lambda}\right)
\end{equation}
where $f$ is a positive function and $\Lambda$ is a cut-off scale. It turns
out that the form of the function $f$ is not important at scales below the
Planck scale. At these scales, the action can be computed using the method of
heat-kernel expansion. For the Dirac operator associated with the
noncommutative geometry of the Standard Model, this computation yields
\cite{AC}
\begin{align}
S_{\mathrm{b}} &  =\frac{48}{\pi^{2}}f_{4}\Lambda^{4}%
{\displaystyle\int}
d^{4}x\sqrt{g}\\
&  -\frac{4}{\pi^{2}}f_{2}\Lambda^{2}%
{\displaystyle\int}
d^{4}x\sqrt{g}\left(  R+\frac{1}{2}a\overline{H}H+\frac{1}{4}c\sigma
^{2}\right)  \nonumber\\
&  +\frac{1}{2\pi^{2}}f_{0}%
{\displaystyle\int}
d^{4}x\sqrt{g}\left[  \frac{1}{30}\left(  -18C_{\mu\nu\rho\sigma}%
^{2}+11R^{\ast}R^{\ast}\right)  +\frac{5}{3}g_{1}^{2}B_{\mu\nu}^{2}+g_{2}%
^{2}\left(  W_{\mu\nu}^{\alpha}\right)  ^{2}+g_{3}^{2}\left(  V_{\mu\nu}%
^{m}\right)  ^{2}\right.  \nonumber\\
&  \qquad\left.  +\frac{1}{6}aR\overline{H}H+b\left(  \overline{H}H\right)
^{2}+a\left\vert \nabla_{\mu}H_{a}\right\vert ^{2}+2e\overline{H}H\,\sigma
^{2}+\frac{1}{2}d\,\sigma^{4}+\frac{1}{12}cR\sigma^{2}+\frac{1}{2}c\left(
\partial_{\mu}\sigma\right)  ^{2}\right]  \nonumber\\
&  +\cdots\nonumber
\end{align}
where $a,c,d,e$ are defined in terms of the Yukawa couplings, $f_{0}=f\left(
0\right)  $ and $f_{k}$ are the Mellin transforms of the function $f$%
\[
f_{k}=%
{\displaystyle\int\limits_{0}^{\infty}}
f\left(  v\right)  v^{k-1}dv,\qquad k>0.
\]
This action contains all bosonic interactions including gravity, gauge
symmetries and those of the Higgs field and a scalar singlet. All couplings
are related at unification scale. The only scale in the expansion is the
cut-off scale $\Lambda.$ The zeroth order term in the heat kernel expansion
gives the cosmological constant, while the first order term gives the
Einstein-Hilbert action and the scalar masses. The second order term, which is
conformal invariant, gives the Yang-Mills and scalar kinetic terms as well as
the second order in curvature terms. A new feature in the spectral Standard
Model is the existence of the singlet field $\sigma$ whose vev gives mass to
the right-handed neutrino. In the Standard Model the Higgs coupling will
become negative at some high energy scale of the order of $10^{11}$ Gev.
Remarkably, this field $\sigma$ plays a very important role in stabilizing the
Higgs coupling which will not become negative at very high energies as well as
being consistent with a low Higgs mass of $125$ Gev \cite{Reselience},
\cite{Karimi}. The form of the gauge and Higgs kinetic terms and potential
implies unification of the gauge couplings and the Higgs coupling. The
universality of the Higgs couplings to all fermion masses implies a relation
between the fermion masses and the gauge vectors masses. A study of the
renormalization group equations shows that these relations are consistent with
present experimental data and predict the top quark mass to be around $170$
Gev. However, gauge couplings do approach each other, but do not meet at one
scale, and thus form a triangle in the energy range $10^{14}-10^{17}$ Gev. At
present these couplings are measured to very high precision, and the non
meeting of the couplings can be taken as a strong indication of the need for
new physics beyond the Standard Model. We will address this point later.

It is then clear from the above results that the simple assumption that
space-time is a noncommutative space which is a product of a continuous
four-dimensional Riemannian spin-manifold tensored with a finite space of $KO$
dimension $6$ gives an excellent geometric explanation of the intricate
details of the Standard Model. We are able to explain and answer many open
questions which are not answered simultaneously in other approaches:

\begin{enumerate}
\item Why there are $16$ fermions per generation.

\item The gauge symmetry $U(1)\times SU\left(  2\right)  \times SU\left(
3\right)  $ and gauge fields as fluctuations of the Dirac operator along
continuous directions.

\item Origin of the Higgs doublet field as fluctuations of the Dirac operator
along discrete directions.

\item Why spontaneous symmetry breaking occurs.

\item Smallness of the neutrino masses due to see-saw mechanism.

\item Stabilizing of the Higgs coupling at large scales.

\item Prediction of the top quark mass.

\item Unification of all fundamental forces including gravity.
\end{enumerate}

These results are quite impressive, especially since there are no alternative
explanations capable of producing them simultaneously. This shows that
adopting the framework of noncommutative geometry for unification of all
fundamental interactions is the right approach. On the other hand, it is also
clear that this is not the final answer because there are still many
unanswered questions, such as why there are only three generations, and the
reason for the hierarchy of the fermions masses. We have also seen that the
gauge coupling constants do not quite meet at a unified scale. This indicates
that there must be new physics beyond the Standard Model. In our
classification of noncommutative spaces, at one point, we made a simplifying
assumption that the connection arising from inner fluctuations of the Dirac
operator belongs to the algebra $\mathcal{A}$ but not to its commutant
$\mathcal{A}^{o},$ which in turn required the Dirac operator to satisfy the
order one condition. This condition need not be satisfied at high energies,
and in seeking to find out whether the Standard Model is only an approximation
of a more unified theory, such possibility must be explored. This brings us
back to the finite algebra taken to be $\mathcal{A}_{F}=\left(  \mathbb{H}%
_{R}\mathbb{\oplus H}_{L}\right)  \oplus M_{4}\left(  \mathbb{C}\right)  .$
Inner fluctuations of the Dirac operator would then yield the connection
\begin{equation}
A=A_{\left(  1\right)  }+JA_{(1)}J^{-1}+A_{\left(  2\right)  },
\end{equation}
where
\begin{align}
A_{\left(  1\right)  } &  =%
{\displaystyle\sum}
a\left[  D,b\right]  \\
A_{\left(  2\right)  } &  =%
{\displaystyle\sum}
\widehat{a}\left[  A_{\left(  1\right)  },\widehat{b}\right]  ,
\end{align}
which satisfies $JA_{\left(  2\right)  }J^{-1}=A_{\left(  2\right)  }.$ The
connection $A_{\left(  2\right)  }$ is not linear, and corresponds to scalar
fields along the off-diagonal components and are in mixed representations with
respect to the gauge symmetry $SU\left(  2\right)  _{R}\times SU\left(
2\right)  _{L}\times SU\left(  4\right)  .$ The gauge vector fields correspond
to inner fluctuations along continuous directions and are those of the
Pati-Salam model where the symmetry group $SU\left(  4\right)  $ assigns the
lepton number as the fourth color \cite{CCS2}. This symmetry then achieves the
unification of leptons and quarks. Depending on the form of the initial Dirac
operator before fluctuations, the most general representation of the Higgs
scalar fields is given by \cite{CCS2}
\begin{align*}
\Sigma_{aI}^{\overset{.}{b}J} &  =\left(  \overline{2}_{R},2_{L},1\right)
+\left(  \overline{2}_{R},2_{L},15\right)  \\
H_{aIbJ} &  =\left(  1_{R},1_{L},6\right)  +\left(  1_{R},3_{L},10\right)  \\
H_{\overset{.}{a}I\overset{.}{b}J} &  =\left(  1_{R},1_{L},6\right)  +\left(
3_{R},1_{L},10\right)  .
\end{align*}
The other possibilities are special cases of the above configuration, obtained
by truncations. The Standard Model is also a special case where $\Sigma
_{aI}^{\overset{.}{b}J}$ is expressed in terms of the Higgs doublet $H$ and
the field $H_{\overset{.}{a}I\overset{.}{b}J}$ is expressed in terms of
$\sigma$ only. Analysis of the renormalization group equations for the
Pati-Salam models show that it is possible to get gauge couplings unification
and this is achieved at energy scales of the order $10^{16}$ Gev
\cite{Walter}. It is then very gratifying to know that general classification
of finite noncommutative spaces of $KO$ dimension $6$ with only one extra
assumption made on one of the algebras to have antilinear isometry symmetry,
picks \ uniquely a Pati-Salam model or one of its truncations, including the
Standard Model. This result is a major step in geometrizing the unification
program and realizing Einstein's dream of finding the right framework for
obtaining a geometric theory where all interactions are dictated by metric
fluctuations, which are now defined through the Dirac operator.

\section{Volume Quantization and Quanta of Geometry}

In the classification given above, the algebra $\mathcal{A}_{F}=\left(
\mathbb{H}_{R}\mathbb{\oplus H}_{L}\right)  \oplus M_{4}\left(  \mathbb{C}%
\right)  $ was the first possibility out of many of the form $\mathcal{A}%
_{F}=\left(  M_{n}\left(  \mathbb{H}\right)  _{R}\mathbb{\oplus}M_{n}\left(
\mathbb{H}\right)  _{L}\right)  \oplus M_{4n}\left(  \mathbb{C}\right)  $, in
addition to making the ad hoc assumption about the antilinear isometry that
reduced the algebra $M_{4n}\left(  \mathbb{C}\right)  $ to $\left(
M_{n}\left(  \mathbb{H}\right)  _{R}\mathbb{\oplus}M_{n}\left(  \mathbb{H}%
\right)  _{L}\right)  $ \cite{CC2}. It is then natural to try to derive the
above results in a unique way without the need of making few assumptions. We
will now do this adopting a different strategy. We start with the observation
that at very small distances, of the order of Planck length, we expect the
discrete nature of space-time to manifest itself in the form of having a
minimal distance. At present the volume of the universe is $\sim10^{60}$ in
Planck units \cite{CC}. In the noncommutative approach, and because of the
need to have compact spectrum for the Dirac operator, space-time is taken to
be of Euclidean signature. We therefore expect the volume of the
four-dimensional space-time to be quantized as integer multiple of the volume
of a unit Planck sphere. Such quantum number is present in the study of maps
from the four-manifold $M_{4}$ to the four-sphere $S^{4}$ in the form of
winding number. We, therefore, must identify the $4-$volume form with the
$4-$form constructed from the $5$ constrained coordinates on the sphere
$Y^{A}$ such that $Y^{A}Y^{A}=1,$ $A=1,\cdots,5$%
\[
\omega=\sqrt{g}dx^{1}\wedge\cdots\wedge dx^{4}=\frac{1}{4!}\epsilon
_{ABCDE}Y^{A}dY^{B}\wedge\cdots\wedge dY^{E}%
\]
This equation forces the volume of the four manifold to be an integer multiple
of the unit sphere $S^{4}$ \cite{Greub}. It turns out, however, that this
condition restricts the topology of $M_{4}$ because the volume form $\omega$
would then not vanish anywhere, and thus the pullback of the maps $Y$ are
non-singular \cite{Moser}. This cannot be always satisfied because the sphere
is simply connected. We will see shortly how \ in noncommutative geometry this
unacceptable restriction is avoided. We first have to write the volume
quantization condition in an index free notation, using the noncommutative
data. In analogy with the Dirac operator which is constructed by tensoring the
connection with the Clifford algebra, the coordinates $Y^{A}$ are tensored  by
Dirac gamma matrices so that $Y=Y^{A}\Gamma_{A}$ where
\begin{equation}
\Gamma_{A}\in C_{\kappa},\quad\left\{  \Gamma_{A},\Gamma_{B}\right\}
=2\kappa\,\delta_{AB},\ (\Gamma_{A})^{\ast}=\kappa\Gamma_{A}.
\end{equation}
Here $\kappa=\pm1$ and $C_{+}=M_{2}\left(  \mathbb{H}\right)  \oplus
M_{2}\left(  \mathbb{H}\right)  $ while $C_{-}=M_{4}\left(  \mathbb{C}\right)
$ \cite{LM}. Since we will be dealing with irreducible representations we take
$C_{+}=M_{2}\left(  \mathbb{H}\right)  .$ We then have
\begin{equation}
Y^{2}=\kappa,\qquad Y^{\ast}=\kappa Y.\label{sphere}%
\end{equation}
and the volume quantization condition takes the simple form \cite{CCM}%
\begin{equation}
\frac{1}{2^{2}(4!)}\left\langle Y\left[  D,Y\right]  ^{4}\right\rangle
=\sqrt{\kappa}\gamma,\label{volume}%
\end{equation}
where $\left\langle \text{ }\right\rangle $ means taking the trace over the
Clifford algebra. Notice that the two conditions in equation (\ref{sphere})
can be combined into one relation $Y^{4}=1.$ Our first observation is that
condition (\ref{volume}) involves the commutator of the Dirac operator $D$ and
the coordinates $Y.$ In momentum space $D$ is the Feynman-slashed
$\slashed{p}$$=\gamma^{\mu}p_{\mu}$ momentum and $Y$ are the Feynman-slashed
coordinates. This suggests that the quantization condition is a higher form of
Heisenberg commutation relation quantizing the phase space formed by
coordinates and momenta. We first notice that although the quantization
condition is given in terms of the noncommutative data, the operator $J$ is
the only one missing. We therefore modify the condition to take $J$ into
account. We first define the projection operator $e=\frac{1}{2}\left(
1+Y\right)  $ satisfying $e^{2}=e$ \ \cite{Connes} but now there are two
possibilities, $Y$ corresponding to the case $\kappa=1$ and $Y^{\prime}$ to
the case $\kappa=-1$. Thus let $Y=Y^{A}\Gamma_{A}$ and $Y^{\prime}=iJYJ^{-1}$
and $\Gamma_{A}^{\prime}=iJ\Gamma_{A}J^{-1}$ so that we can write
\begin{equation}
Y=Y^{A}\Gamma_{A},\qquad Y^{\prime}=Y^{\prime A}\Gamma_{A}^{\prime},
\end{equation}
satisfying $Y^{2}=1$ and $Y^{\prime2}=1.$ The projection operators $e=\frac
{1}{2}\left(  1+Y\right)  $ and $e^{\prime}=\frac{1}{2}\left(  1+Y^{\prime
}\right)  $ satisfy $e^{2}=e$, $e^{\prime2}=e^{\prime}$ with $e$ and
$e^{\prime}$ commuting. This allows to define the projection operator
$E=ee^{\prime}$ and the associated field
\begin{equation}
Z=2E-1,
\end{equation}
satisfying $Z^{2}=1.$ The conjectured quantization condition takes the elegant
form of a two-sided relation \cite{CCM1}, \cite{CCM}%

\begin{equation}
\left\langle Z\left[  D,Z\right]  ^{4}\ \right\rangle =\gamma.\label{SM}%
\end{equation}
One of the remarkable properties of four dimensions is that the quantization
condition in terms of the $Z$ coordinates splits into the sum of two pieces,
one in terms of $Y$ and the other in terms of $Y^{\prime}$%
\begin{equation}
\left\langle Z\left[  D,Z\right]  ^{4}\right\rangle =\left\langle Y\left[
D,Y\right]  ^{4}\right\rangle +\left\langle Y^{\prime}\left[  D,Y^{\prime
}\right]  ^{4}\right\rangle .
\end{equation}
This property is only shared with dimension $2$ but not with higher dimensions
where interference terms do arise. This implies that the volume form of the
$4-$dimensional Riemannian manifold is the sum of two $4-$forms and thus
\begin{equation}
\omega=\frac{1}{4!}\epsilon_{ABCDE}\left(  Y^{A}dY^{B}\wedge\cdots\wedge
dY^{E}+Y^{^{\prime}A}dY^{^{\prime}B}\wedge\cdots\wedge dY^{^{\prime}E}\right)
\label{integer}%
\end{equation}
Thus the volume of the four-manifold (in multiples of unit Planck spheres)  is
the sum of two integers, the winding numbers associated with the two maps $Y$
and $Y^{\prime}.$ The restriction on the topology of $M_{4}$ to be that of a
sphere is now removed because the sum of the pullbacks of $Y$ and $Y^{\prime}$
does not vanish anywhere and thus each one of them could vanish separately,
but not simultaneously. We have shown that for a compact connected smooth
oriented manifold with $n\leq4$ one can find two maps $\phi_{+}^{\#}\left(
\alpha\right)  $ and $\phi_{-}^{\#}\left(  \alpha\right)  $ whose sum does not
vanish anywhere, satisfying equation (\ref{integer}) such that $%
{\displaystyle\int\limits_{M}}
\omega\in\mathbb{Z}.$ The proof for $n=4$ is more difficult and there is an
obstruction unless the second Stieffel-Whitney class $w_{2}$ vanishes, which
is satisfied if $M$ is required to be a spin-manifold and the volume to be
larger than or equal to five units \cite{CCM}, \cite{CCM1}. 

We therefore, take as our starting point the higher form of Heisenberg
commutation relations with the field $Z$ defining two separate maps from
$M_{4}$ to the sphere $S^{4}$
\[
Z=\frac{1}{2}\left(  Y+1\right)  \left(  Y^{\prime}+1\right)  -1,\qquad Y\in
M_{2}\left(  \mathbb{H}\right)  ,\quad Y^{\prime}\in M_{4}\left(
\mathbb{C}\right)
\]
and belonging to the finite algebra%
\[
\mathcal{A}_{F}=M_{2}\left(  \mathbb{H}\right)  \oplus M_{4}\left(
\mathbb{C}\right)  ,
\]
However, the maps $Y$ and $Y^{\prime}$ are functions of the coordinates of the
manifold $M_{4}$ and therefore the algebra associated with this space must be
\begin{equation}
\mathcal{A}=C^{\infty}\left(  M,\mathcal{A}_{F}\right)  =C^{\infty}\left(
M\right)  \otimes\mathcal{A}_{F}.
\end{equation}
The associated Hilbert space is
\begin{equation}
\mathcal{H}=L^{2}\left(  M_{4},S\right)  \otimes\mathcal{H}_{F}.
\end{equation}
The Dirac operator mixes the finite space and the continuous manifold
non-trivially%
\begin{equation}
D=D_{M}\otimes1+\gamma_{5}\otimes D_{F},
\end{equation}
where $D_{F\text{ }}$ is a self adjoint operator in the finite space. The
chirality operator is $\gamma=\gamma_{5}\otimes\gamma_{F},$  and the
anti-unitary operator $J$ is given by $J=J_{M}\gamma_{5}\otimes J_{F}.$ At
this point we realize that we have obtained exactly the same noncommutative
space obtained before and all subsequent analysis becomes identical.

\section{\bigskip Conclusions}

We have thus identified the noncommutative space that corresponds to the
structure of space-time. We started first by classifying all possible
noncommutative spaces which are tensor products of continuous four-dimensional
space, times finite spaces with $KO$ dominions $6,$  as dictated by the
absence of mirror fermions. We showed that the algebra of the finite space is
the sum of two matrix algebras. We then assumed that one of the algebras
satisfy a certain antilinear isometry. This yielded a class of spaces, the
first of which is the one that produces the Standard Model with all its
intricacies. The result comes out in a unique way by assuming a differential
condition on the Dirac operator where the connection obtained by inner
fluctuations is restricted to the algebra but not its commutant and is linear.
The Standard Model will hold as a unified model up to very high energies,
however, it is not adequate at energies higher than $10^{14}$ Gev as evidenced
from the failure of the gauge coupling constants to unify. This suggests that
at very high energies the order one condition on the Dirac operator be
removed. In an alternative approach, we started with an axiom representing a
higher Heisenberg commutation relation quantizing the phase space of Dirac
operators and coordinates provided by the two \ maps from the manifold to four
spheres. The advantage of the second approach, is that there is no need to
make various assumptions, physical, or ad hoc. This axiom of quantization is
enough to determine fully the noncommutative space defining the structure of
space-time. We have the physically satisfying result that the volume of
space-time is quantized in terms of two kinds of quanta associated with the
two maps $Y$ and $Y^{\prime}.$ Inner fluctuations of the Dirac operator over
the algebra $\mathcal{A}$ then results in gauge and Higgs fields of the
Pati-Salam model with the symmetry of $SU\left(  2\right)  _{R}\times
SU\left(  2\right)  _{L}\times SU\left(  4\right)  .$ The $16$ fermions of
each family are in the fundamental representation of the Hilbert space. The
Standard Model is a special truncation of the Pati-Salam model. We have thus
presented very strong evidence that noncommutative geometry is the correct
framework to define the structure of space-time. We have determined the
precise noncommutative space that reproduces all known particle interactions
including gravity. Confidence in these results is strengthened by the fact
that two different strategies, although completely not correlated, give the
same answer. The advantage of the second approach is that in this case the
noncommutative space is obtained uniquely without the need to make more
assumptions or restrict ourselves to the simplest possibility. This shows that
the idea of volume quantization is a fundamental one, and not only determine
the noncommutative space defining space-time, but also defines the two
different quanta needed to construct the four dimensional manifold.

We have, therefore succeeded in removing many of the mysteries associated with
the Standard Model and pointed the way to the physics beyond.  We have also
answered many of the questions that defied explanations for so long.
Naturally, many more questions remain, and it is a big challenge to extend the
ideas presented here to uncover the remaining mysteries.

\begin{acknowledgement}
I would like to thank Alain Connes for a fruitful collaboration on the topic
of noncommutative geometry since 1996. I would also like to thank Walter van
Suijlekom and Slava Mukhanov for essential contributions to this program of
research. This research is supported in part by the National Science
Foundation under Grant No. Phys-1518371.
\end{acknowledgement}


\begin{thebibliography}{99}                                                                                               %
\bibitem {Connesbook}A. Connes, \textit{Noncommutative Geometry}. Academic
Press, 1994.

\bibitem {Reality}A. Connes, \textit{Noncommutative Geometry and Reality, }J.
Math. Phys. \textbf{36 }(1995) 6194.

\bibitem {Connes}A. Connes, A Short Survey of Noncommutative Geometry, arXiv: hep-th/0003006.

\bibitem {Gracia}J. Gracia-Bondia, J. Varilly and H. Figueroa,
\textit{Elements of Noncommutative Geometry, Birkhauser. }

\bibitem {Kac}Mark Kac, \textit{Can One Hear the Shape of a Drum? }The
American Mathematical Monthly, \textbf{73 (}1996) 1-23.

\bibitem {Milnor}J. Milnor, \textit{Eigenvalues of the Laplac Operator on
Certain Manifolds, }Proc. Nat. Acad. Sci. USA, \textbf{51} (1964) 542.

\bibitem {AC2}A. H. Chamseddine and A. Connes, \textit{The Spectral Action
Principle, }Comm. Math. Phys. \textbf{186 }(1997) 731.

\bibitem {saddam}A. H. Chamseddine, A. Connes and M. Marcolli, \textit{Gravity
and the Standard Model with Neutrino mixing, }Adv.Theo.Math. Phys. \textbf{11
}(2007) 991-1089.

\bibitem {AC}A. H. Chamseddine and A. Connes, \textit{Noncommutative Geometry
as a framework for unification of all fundamental interactions including
gravity, }Fortsch. Phys. \textbf{58 }(2010) 553. 

\bibitem {Gros}A. H. Chamseddine, A. Connes, \textit{Space-Time from the
Spectral Point of View, }Proceedings 12$^{\text{th}}$ Marcel Grossmann
Meeting, Paris, July 12-18 2009, Editors, T. Damour et al.

\bibitem {CC2}A. H. Chamseddine and A. Connes, \textit{Why the Standard Model}
, J. Geom. Phys. \textbf{58}, (2008) 38.

\bibitem {CCS}A. H. Chamseddine, A. Connes, W. van Suijlekom, \textit{Inner
Fluctuations in Noncommutative Geometry without the First Order Condition,
}Jour. Geom. Phys. \textbf{73 }(2013) 222.

\bibitem {Reselience}A. H. Chamseddine and A. Connes, \textit{Reselience of
the Spectral Standard Model, }JHEP \textbf{09 }(2009) 104.

\bibitem {Karimi}H. Karimi, \textit{Implications of the Complex Singlet Field
for Noncommutative Geometry Model, }JHEP \textbf{1712} (2017) 040.

\bibitem {CCS2}A. H. Chamseddine, A. Connes, W. van Suijlekom, \textit{Beyond
the Spectral Standard Model: Emergence of Pati-Salam Unification} , JHEP
\textbf{11 }(2013) 132.

\bibitem {Walter}A. H. Chamseddine, A. Connes, W. van Suijlekom, \textit{Grand
Unification in the Spectral Pati-Salam Model, }JHEP \textbf{1511 }(2015) 011.

\bibitem {CC}A. H. Chamseddine and A. Connes, \textit{The Uncanny Precision of
the Spectral Action} , Comm. Math. Phys. \textbf{293}, (2010) 867-897.

\bibitem {Greub}W. Greub, S. Halperin and R. Vanstone, \textit{Connections,
Curvature and Cohomology, }volumes 1-3, and in particular pages 347-351 volume
2 (sphere maps), Academic Press (1973).

\bibitem {LM}H. Lawson and M. Michelson, \textit{Spin Geometry, }Princeton
University Press, 1989.

\bibitem {Moser}J. Moser, \textit{ On the volume elements on a manifold}.
Trans. Amer. Math. Soc. 120 (1965) 286-294.

\bibitem {CCM1}A. H. Chamseddine, A. Connes, Viatcheslav Mukhanov,
\textit{Geometry and the Quantum:Basics, arXiv:1411.0977, JHEP }\textbf{12
}(2014) 098.

\bibitem {CCM}A. H. Chamseddine, A. Connes, V. Mukhanov, \textit{Quanta of
Geometry: Noncommutative Aspects, }Phys. Rev. Lett. \textbf{114 }(2015) 091302.
\end{thebibliography}
\end{document}